\documentclass[a4paper,11pt]{article}

\usepackage{jcappub} 
\usepackage{amsmath}

\usepackage[T1]{fontenc} 

\title{\boldmath On the cosmology of Scalar-tensor-vector gravity theory}

\author[a]{Sara Jamali,}
\author[a]{Mahmood Roshan}
\author[b]{and Luca Amendola}

\affiliation[a]{Department of Physics, Ferdowsi University of Mashhad,\\P.O. Box 1436, Mashhad, Iran}
\affiliation[b]{Institute for Theoretical Physics, University of Heidelberg,\\ Philosophenweg 16, D-69120 Heidelberg, Germany}

\emailAdd{sara.jamali@mail.um.ac.ir}
\emailAdd{mroshan@um.ac.ir}
\emailAdd{l.amendola@thphys.uni-heidelberg.de}

\abstract{We consider the cosmological consequences of a special scalar-tensor-vector theory of gravity, known as  MOG  (for MOdified Gravity), proposed to address the dark matter problem. This theory introduces two scalar fields $G(x)$ and $\mu(x)$, and one vector field $\phi_{\alpha}(x)$, in addition to the metric tensor. We set the corresponding self-interaction potentials to zero, as in the standard form of MOG. Then using the phase space analysis in the flat Friedmann-Robertson-Walker background, we show that the theory possesses a viable sequence of cosmological epochs with acceptable time dependency for the cosmic scale factor. We also investigate MOG's potential as a dark energy model and show that extra fields in MOG cannot provide a late time accelerated expansion. Furthermore, using a dynamical system approach to solve the non-linear field equations numerically, we calculate the angular size of the sound horizon, i.e. $\theta_{\text{s}}$, in MOG. We find that $8\times 10^{-3}\text{rad}<\theta_{\text{s}}<8.2\times 10^{-3} \text{rad} $ which is way outside  the current observational bounds. Finally, we generalize MOG to a modified form called mMOG, and we find  that mMOG passes the sound-horizon constraint. However, mMOG also cannot be considered as a dark energy model un;ess one adds a cosmological constant, and more importantly, the matter dominated era is still slightly different from the standard case.}

\begin{document}
\maketitle
\flushbottom

\section{Introduction}
\label{intro}
MOG (MOdified Gravity) is a scalar-tensor-vector theory of gravity proposed by J. W. Moffat that introduces two scalar fields and one Proca vector field besides the metric tensor in the action \cite{moffat2}. The vector field is directly coupled to ordinary matter. This theory has been introduced as a solution to the dark matter problem. In other words in this theory there is no dark matter particle and the dark matter problem is addressed by changing the gravity sector. From this perspective one may compare MOG to modified Newtonian dynamics (MOND) \cite{Milgrom:1983ca} and its relativistic version TeVeS \cite{Bekenstein:2004ne}. MOG has been widely investigated in the literature in order to check its viability to pass the astrophysical tests related to the dark matter problem. For example by fitting appropriate values for the free parameters of the theory , i.e. $\alpha$ and $\mu$, in the weak field limit, MOG explains the flat rotation curve of galaxies without invoking cold dark matter particles \cite{Moffat:2013sja}. Tests of rotational velocity curve in the Milky Way shows that MOG can fit the data better than MOND \cite{Moffat:2014pia}.  The mass discrepancy in the galaxy clusters is also explained by MOG in \cite{Brownstein:2005dr, Moffat:2013uaa}. The global stability of disk galaxies in the context of MOG has been studied in \cite{Ghafourian:2017wfr} using N-body simulations, where it has been shown that disk galaxies are more stable in MOG than in Newtonian gravity. More specifically, like rigid dark matter halos, MOG can stabilize the disk against bar instability. For an analytic study of the stability of the Maclaurin disk in MOG we refer the reader to \cite{Roshan:2016ygw}. The weak field limit of the theory and the local stability of the disks against small perturbations has been studied in  \cite{ra} and \cite{Roshan:2015gra} respectively. The test particle equations of motion for spinning and non-spinning test particles in MOG has been investigated in \cite{mn}. Finally,  it has been claimed in \cite{Israel:2016qsf} that MOG has the potential to explain merging cluster dynamics (e.g. the Bullet Cluster and the Train Wreck Cluster) without assuming a dark matter component for the clusters.

Although the cosmological behaviour of this theory has been considered in some papers \cite{Roshan:2015uta,Moffat:2014bfa,Moffat:2015bdaf}, we believe there are several aspects which require more investigation. For example in \cite{Jamali:2016zww} it has been shown that for a model of MOG, which is different from the standard version of the theory, there is no standard matter dominated phase in the cosmic history of the model. In other words, although some models of MOG are successful in addressing the dark matter problem in the galaxy and galaxy cluster scales, they may not work properly in the cosmological scale.

In this paper we investigate first the standard version of MOG and check its cosmological behaviour using the phase space analysis. This numerical method reveals the main features of the model without solving the modified Friedmann equations for cosmic scale factor $a(t)$, for example see \cite{Amendola:1999qq,Amendola:2006kh, Amendola:2007nt}.  More specifically, with this method one may easily check if the model has an appropriate sequence of the standard cosmological epochs. Naturally, we expect the evolution of universe starts from a radiation dominated epoch, continues toward a standard matter dominated era that allows the structure formation and ends at the stable accelerated dark energy dominated phase.

After investigating the existence of viable cosmological epochs in MOG, we check its consistency with another precise cosmological observation, namely the angular size of the sound horizon  derived form Cosmic Microwave Background observations. In order to make MOG compatible with this observational constraint, we introduce a modified version of MOG, called mMOG, and check the viability of mMOG as a dark energy theory, as well.

This paper is organized as follows. In section \ref{sec-1} we briefly introduce MOG and its modified Friedmann equations. Then, in section \ref{sec-2}, we introduce the dynamical system variables and rewrite the modified Friedmann equations with respect to dimensionless variables of the phase space. We also determine the relevant fixed points and investigate their stability and physical interpretations. In sections \ref{sec-4.1} and \ref{sec-4.2} we answer some important questions about MOG. In fact, we investigate if the theory could work properly in the absence of the nonzero component of the vector field, i.e. $\phi_0$. Also the question of whether this theory could play the role of dark energy, as well as the dark matter, is studied in this section. In fact, we analyse the system in both cases of existence and absence of $\phi_0$, while $\Lambda$ is zero. A modified version of MOG (mMOG) is also presented in this section. The angular size of the sound horizon $\theta_{\text{s}}$ is investigated in section \ref{sec-3.1} for both the standard and the modified version of MOG. We compare the result with the relevant observation and show that $\theta_{\text{s}}$ in the standard version of MOG is much below the observed value while we  find a modified version  which passes this observational constraint. Finally, conclusions are drawn in sec \ref{sec-5}.

\section{Modified Friedmann equations in MOG}\label{sec-1}
Let us start with the action of MOG\footnote{to be denoted, in case of confusion, standard MOG.}. We use the exact form introduced in \cite{Moffat:2007nj}
\begin{eqnarray}
\!\!S\!\!=\!\!\int \!\! \sqrt{-g} d^{4}x \bigg[
 \frac{(R\!+\!2\Lambda)}{16 \pi G}  \!\!+\!\!\frac{\omega_{0}}{4 \pi}\left(\frac{1}{4} B_{\mu \nu}B^{\mu \nu} \!\!+\!\!V_{\phi}\right)
 \!\!+\!\!\frac{1}{2 G}\left(\frac{\nabla_{\mu} G \nabla^{\mu} G}{G^2}\!+\!\frac{\nabla_{\nu} \mu \nabla^{\nu} \mu}{\mu^2}\right)\bigg] \!\!+\!\!S_{\text{M}}
\label{action}
\end{eqnarray}
where $R$ is the Ricci scalar, $\Lambda$ is
a positive constant corresponding to the cosmological constant in the Einstein-Hilbert action, $\omega_{0}$ denotes a positive constant and $B_{\mu \nu}=\nabla _{\mu} \phi_{\nu}-\nabla_{\nu}\phi_{\mu}$ is an anti-symmetric tensor. The potential $V_{\phi}$ is chosen as $V_{\phi} =- \frac{1}{2} \mu^2 \phi_{\alpha} \phi^{\alpha} $; this  reduces to $-\frac{1}{2} \mu^2 \phi_{0}^2 $, in a homogeneous and isotropic cosmology where $\phi_{0}$ is the zeroth component of the vector field. Furthermore $G$ and $\mu$ are scalar fields and $S_{\text{M}}$ is the matter action. As mentioned,  $B_{\mu \nu}$ is an anti-symmetric tensor and consequently it vanishes in the cosmological context. In this way, we absorb the parameter $\omega_0$ in the definition of $\mu$ as $\sqrt{\omega_0} \mu \rightarrow  \mu$. Moreover, $\omega_{0}$ in $\frac{\omega_{0}}{4 \pi}V_{\phi}$ is also absorbed by $V_{\phi}$ and it will not appear in the rest of our calculations.

Varying the action with respect to the fields, one can find the corresponding field equations. In order to study the cosmological consequences of MOG, we assume a flat Friedmann-Robertson-Walker (FRW) metric 
\begin{equation}
ds^2=dt^2-a(t)^2(dx^2+dy^2+dz^2)
\end{equation}

Substituting the metric in the field equations we find the following modified Friedmann equations, which are exactly the equations already obtained in \cite{Moffat:2007nj},
\begin{eqnarray}
\begin{split}
&H^2 =\frac{8 \pi G}{3} \rho + \frac{\Lambda}{3} + \frac{\dot{G}\dot{a}}{G a} 
 - \frac{4 \pi}{3}( \frac{\dot{\mu}^2}{\mu^2} +  \frac{\dot{G}^2}{G^2})+ \frac{1}{3} G \mu^2 \phi_0^2\\&
\frac{\ddot{a}}{a}=-\frac{4\pi G}{3}(\rho+3 p)\!+\!\frac{1}{2}\frac{\ddot{G}}{G}+(\frac{8\pi -3}{3})\frac{\dot{G}^2}{G^2}-\frac{2 G \mu^2 \phi_0^2 }{3}
+\frac{\Lambda}{3}+\frac{1}{2}\frac{\dot{G}}{G}H+\frac{8 \pi}{3}\frac{\dot{\mu}^2}{\mu^2}\\&
\frac{\ddot{G}}{G}= \frac{3}{8 \pi}\left(\frac{\ddot{a}}{a}+ H^2 \right)-3 H \frac{\dot{G}}{G} +\frac{3\dot{G}^2}{2 G^2} -\frac{\dot{\mu}^2}{ 2 \mu^2}-\frac{\Lambda}{8 \pi}\\&
\frac{\ddot{\mu}}{\mu}= \frac{\dot{\mu}^2}{\mu^2}-3 H \frac{\dot{\mu}}{\mu}+\frac{\dot{G}}{G}\frac{\dot{\mu}}{\mu}- \frac{1}{4 \pi}G \mu^2 \phi^2\\&
\frac{\partial V_{\phi}}{\partial \phi_0}=16\pi J^0
\end{split}
\label{FE1}
\end{eqnarray}
where a dot denotes derivative with respect to cosmic time $t$, $\rho = \rho_{\text{m}} + \rho_{\text{r}}$  where  $\rho_\text{m}$ and $\rho_{\text{r}}$ are matter and radiation energy densities, respectively, $H(t)=\frac{\dot{a}}{a}$ and $J^{0}$ is the nonzero component of "fifth force" matter current defined as
 \begin{eqnarray}\label{current1}
J^{0}=-\frac{1}{\sqrt{-g}}\frac{ \delta S_\text{M}}{\delta \phi_{0}}
 \end{eqnarray}
 The coupling between matter and the vector field leads to a nonzero fifth-force current. In fact this is a central feature in MOG for addressing the dark matter problem. Naturally, one has to postulate the way by which the vector field is coupled to matter. It is convenient to use $J^{0}=\kappa \rho_{\text{m}}$, where $\kappa$ is a coupling constant \cite{Moffat:2015bda}. It is important to mention that in this version of MOG the continuity equation  holds. Therefore, as in the standard cosmological model, we have $\rho_{\text{m}} \propto a^{-3}$ and $\rho_{\text{r}} \propto a^{-4}$.

It is important to mention that, as it is obvious from the first modified Friedmann equation, i.e. the first equation in \eqref{FE1},  the kinetic terms of scalar fields $\mu$ and $G$ appear with the "wrong" sign with respect to the standard Klein-Gordon scalar field Lagrangian. Therefore,  despite the claims in the relevant literature, for example see \cite{moffat2}, this version of MOG is not ghost free. However, our aim
here is to study just the classical version, and we impliclty assume we deal with  an effective classical limit of a stable, fundamental quantum field theory. In the Appendix \ref{app} we discuss the stability of the theory, and show that it is tachyon-free.  In other words, it turns out that the squared effective mass of the scalar perturbations is positive, and consequently there is no tachyonic instability. Therefore, from the stability point of view, the theory works properly at the classical level.

\section{Phase space analysis of MOG}\label{sec-2}
Phase space analysis, or the dynamical system approach, is a numerical method that has been widely used to investigate the cosmological behaviour of modified theories of gravity. For a brief introduction to this method we refer the reader to \cite{Copeland:2006wr}.  Let us construct a set of autonomous equations from \eqref{FE1} by defining the following dimensionless variables
\begin{eqnarray}
\begin{split}
& y=\frac{8 \pi G}{3 H^2} \rho_{\text{m}},\, r=\frac{8 \pi G}{3 H^2} \rho_{\text{r}},\, z =\frac{\dot{G}}{G H}, \, m = \sqrt{\frac{4 \pi}{3}} \frac{\dot{\mu}}{\mu H},\, x^2 = \frac{\Lambda}{3 H^2}, \,  Q=\frac{G}{3}\left(\frac{16 \pi \kappa\rho_{\text{m}} }{ H \mu }\right)^2
\end{split}
\label{var}
\end{eqnarray}
The cosmic density parameters $\Omega_i$ are related to dimensionless variables as
$$\Omega_{\text{m}}\!\! =\!\! y,~~\Omega_{\text{r}} \!\!= \!\!r,~~\Omega_{\Lambda} \!\!=\!\! x^2,~~\Omega_{\mu} \!\!=\!\! -m^2,~~\Omega_{\text{G}} \!\!=\!\! z-\frac{4\pi}{3} z^2 $$
Substituting the dynamical variables into equations \eqref{FE1} and assuming $G>0$, and consequently $Q>0$, after some algebraic manipulations we find a constraint equation
\begin{eqnarray}
y+r+x^2+z-m^2-\frac{4 \pi}{3}z^2 \leq 1
\end{eqnarray}
and the following autonomous first order differential equations
\begin{equation} \label{zprime}
\begin{split}
&y^{\prime}=y  (6 m^2-8 \pi  \left(4 r+12 x^2+6 y-(z-2) z-6\right)+6 x^2-3 z-3) \\&
r^{\prime} = r \left (6 m^2-8 \pi  \left(4 r+12 x^2+6 y-(z-2) z-4\right)+6 x^2-3 z\right)\\&
x^{\prime} = x  \left(3 m^2-4 \pi  \left(4 r+6 \left(2 x^2+y-2\right)-z^2+4 z\right)+3 \left(x^2-2\right)\right)\\&
z^{\prime} =3  \left(m^2 z-2 m^2+2 r+4 x^2+3 y+3 z-2\right)  \\&~~~~~~~~~~~~~~~~~~~+z \left(3 \left(x^2-z\right)-4 \pi  \left(4 r+12 x^2+6 y-(z-2) z\right)\right)\\&
m^{\prime}= \frac{16 \pi -3}{(96 \pi -18)}\bigg[ -18 m^3+3 \sqrt{\frac{3}{\pi }} (16 \pi -3) m^2+6 m (4 \pi  (4 r+12 x^2+6 y-z^2)\\&~~~~~~~~~~~~~~~~~~~ -3 (x^2-z+1))
+\sqrt{\frac{3}{\pi }} (16 \pi -3) \left(-3 \left(r+x^2+y-1\right)+4 \pi  z^2-3 z\right)\bigg]
\end{split}
\end{equation}
where in these equations, a prime stands for derivative with respect to $\ln a$. Our cosmological model is equivalent to a five dimensional dynamical system. Now it is easy to express the effective equation of state parameter
\begin{equation}\label{lan}
\omega_{\text{eff}}=\frac{p_{\text{tot}}}{\rho_{\text{tot}}} = -1-\frac{2\dot{H}}{3H^2}
\end{equation} 
and the equation of state for dark energy $\omega_{\text{DE}}= \frac{ p_\text{\text{DE}}}{\rho_{\text{DE}}}$ with respect to the dynamical system variables. For more details we refer the reader to \cite{Jamali:2016zww}. Finally, $\omega_{\text{DE}}$  and $\omega_{\text{eff}}$ take the  form
\begin{eqnarray}
\begin{split}
&\omega_{\text{DE}}=\frac{(\!16\! \pi \!  (\pi  (16 \pi\! -\!3)-9)+9) m^2\!+\!3 \left(-3 \beta  r\!+\! 8 \pi  \left(2 (\beta \!+\! 2) r\!+\! \zeta\right)-6 x^2+3\right)}{(16 \pi -3) \left(\left(16 \pi ^2-9\right) m^2+9 \beta  (r+y)-9\right)}\\&
\omega_{\text{eff}}=\frac{6 m^2-8 \pi  \left(4 r+\zeta\right)+6 x^2-3}{48 \pi -9}
\end{split}
\label{oms}
\end{eqnarray}
Where $\beta=G_{\text{N}}/G$ and $\zeta=12 x^2+6 y-(z-4) z-6$. Now let us find the critical points. To do so one should set to zero the right hand side of equations \eqref{zprime} and find the corresponding roots $(x,y,r,z,m)$. Furthermore, by constructing the stability matrix, one may consider the linear stability of the critical points. The result has been summarized in Table \ref{tab1}.  Note that in this paper we keep all numbers up to two digits. Surprisingly, for each cosmological epoch there are two critical points. It is important to mention that, as in the standard $\Lambda$CDM model, there is no free parameter in the fixed points. In the following we discuss the physical meaning of the points.

\begin{table*}[!]\renewcommand{\arraystretch}{1.5}
\small
\begin{center}
\begin{tabular}{|l cccc|}
\hline
Point& $(x,y,r,z,m) $ & ~~~~~~$(\Omega_{\Lambda},\Omega_{\text{m}},\Omega_{\text{r}},\Omega_{\text{G}},\Omega_{\mu}) $~~~~~~ &$\omega_{\text{eff}}$ & Stability\\ \hline \hline 
$f_1$ & $(0,0,1.60,-0.53,-2.05)$&(0,~0,~1.60,~-1.72,~-4.19)& 0.51 & unstable\\ 
$f_2$ & $(0,0,1,0,0)$ &(0,~0,~1,~0,~0)&$\frac13$ & unstable\\ 
$f_3$ & $(0,1.71,0,-0.74,-3.07)$ & (0,~0,~1.60,~-3.03,~-9.42)&0.25 & unstable\\ 
$f_4$ & $(0,0.97,0,0.04,0)$ &(0,~0.97,~0,~0.03,~0)&-0.01 & unstable\\ 
$f_5$ & $(1.85,0,0,-2.06,-8.24)$ & (3.42,~0,~0,~-19.76,~-67.93)&-1.00 & unstable\\ 
$f_6$ & $(0.98,0,0,0.04, 0)$ &(0.97,~0,~0,~0.03,~0)&-1  & stable\\ 
\hline
\end{tabular}
\caption{Fixed points and their stability character}\label{tab1}
\end{center}
\end{table*}

\begin{itemize}
	\item {\textbf{\textit{$f_{1,2}$: Radiation-dominated phases:}}}
\end{itemize}
The point $f_1$ is an unstable $G\mu$-radiation dominated era for which eigenvalues of stability matrix are $(0, 0, 1.60, -0.53, -2.05)$. One should note that existence of even one positive eigenvalue means that the critical point is unstable. However, the value of $\omega_{\text{eff}}$ shows that this epoch is drastically different from the standard radiation dominated epoch. On the other hand, $f_2$ is a pure radiation dominated epoch for which $\omega_{\text{eff}}=\frac13$. In this case eigenvalues are $(2, -2, 1, -1, -1)$ which shows that, as expected, $f_2$ corresponds to an unstable radiation phase. Therefore, one may conclude that MOG possesses a standard radiation dominated epoch in which the scalar fields are constant and do not have a significant contribution.

\begin{itemize}
	\item {\textbf{\textit{$f_{3,4}$: matter dominated eras:}}}
\end{itemize}
$f_3$ is a $G\mu$-matter dominated point.  $(-3.88, 2.02, -1.87, 1.87, -1.01)$ are the eigenvalues of stability matrix that shows $f_3$ is unstable. In this phase the scale factor grows as $a(t) \propto t^{0.53}$ which is much slower than the standard  case and cannot be considered as a standard matter dominated epoch. One can expect the growth of perturbations in this epoch to be very different from the standard case, although this has to be checked numerically. On the other hand, $f_4$ is a $G$-matter dominated point for which $a(t) \propto t^{0.67}$.

 Although this is close to the standard case, the effective equation of state parameter is negative. Therefore in principle it is different from the  standard matter dominated universe. $f_4$ is unstable since the eigenvalues of the stability matrix are $(-2.99,\! 1.48,\! -1.48,\! -1.48,\! -0.99\!)$. We can simply conclude that the standard version of MOG has a matter dominated era that behaves very similar to the standard model. This is not the case in some metric $f(R)$ models \cite{Amendola:2006we}. It should be noted that $f_4$ corresponds to an exact solution for which the scalar field $\mu(t)$ is constant and $G(t)$ varies with time as $G(t) \propto t^{0.03}$.
\begin{itemize}
	\item {\textbf{\textit{$f_{5,6}$: $\Lambda$ dominated solutions:}}}
\end{itemize}
$f_5$ corresponds to an unstable phantom phase.  For this critical point $\omega_{\text{eff}} \simeq -1.00$ and the eigenvalues of the stability matrix are $(-10.82,-6.06,5.76,-5.06,-5.06)$ which does not represent  a stable late time $\Lambda$ dominated solution since there is one positive eigenvalue. So, $f_5$ is not a standard late time solution. This solution cannot  be considered as an early time inflationary fixed point as well because of the existence of negative eigenvalues.  On the other hand, $f_6$ is stable with eigenvalues  $(-5.96, -3.96, -2.96, -2.96, -2.96)$. In this case $\omega_{\text{eff}}= -1 $ which is reminiscent of the standard de Sitter universe. In this era, $G(t) \propto e^{0.04t}$ and $\mu(t)$ is approximately constant.

\begin{figure}[tbp]
\centering 
\includegraphics[width=.49\textwidth,origin=c]{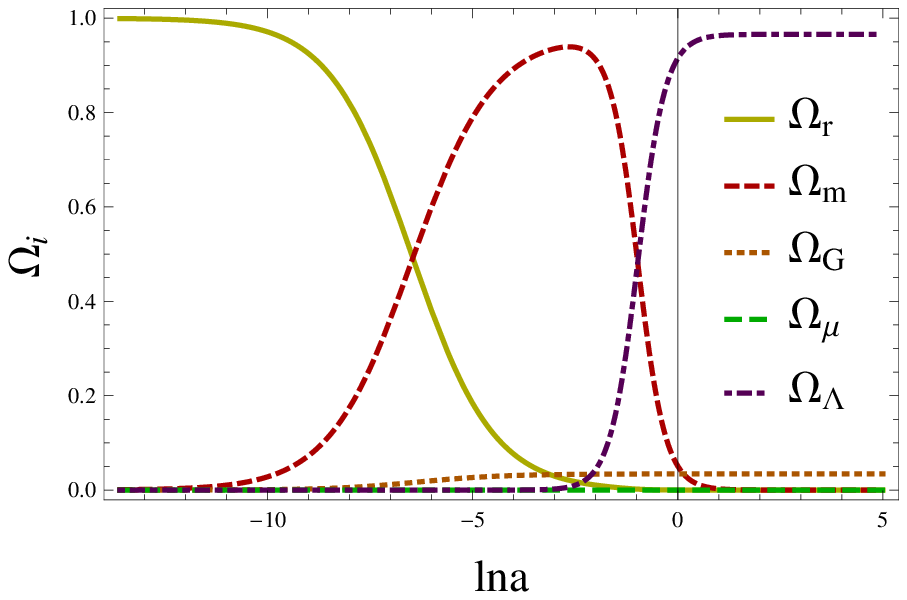}
\includegraphics[width=.49\textwidth,origin=c]{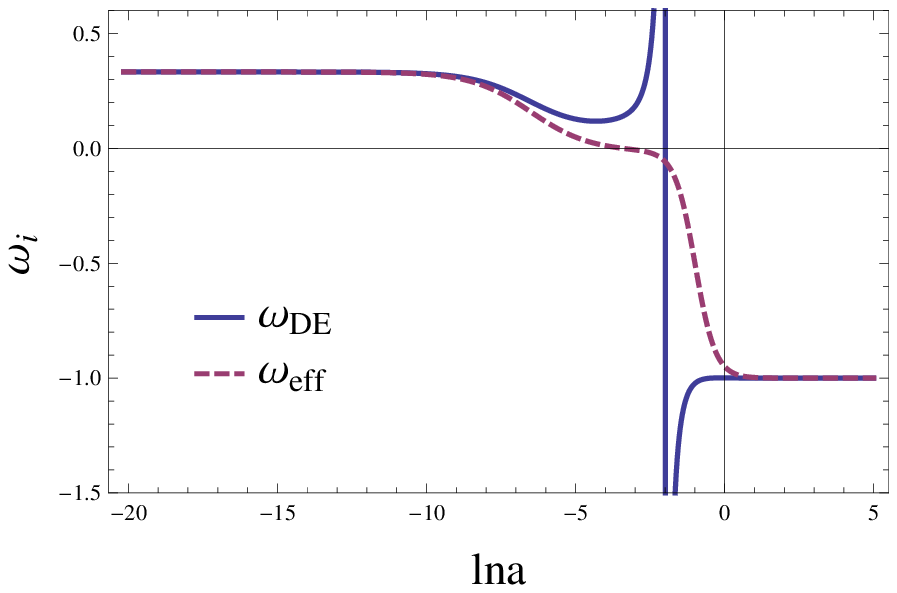}
\caption{\label{Range} \textit{left panel}: The time evolution of the cosmic density parameters $\Omega_\text{i}$ in MOG. The initial conditions are chosen deep in radiation dominated epoch, i.e. $\ln a = -20.4$. The initial conditions are $\Omega_{\text{r}}=0.99$, $\Omega_{\text{m}} =1.18 \times 10^{-6} $, $\Omega_{\text{G}} \approx 3.3 \times 10^{-16}$ and $\Omega_{\mu} =-1.1 \times 10^{-31} $and $\Omega_{\Lambda}=1.6 \times 10^{-31}$. The initial conditions are chosen in a way to lead to $\Omega_{\text{m0}}=0.05$ and $\Omega_{\text{r0}}=8 \times 10^{-5}$. A viable sequence of cosmological epochs is seen. \textit{right panel}: The behaviour of $\omega_{\text{eff}}$ and $\omega_{\text{DE}}$ for the same initial conditions of left panel.}
\end{figure}

A valid cosmological path starts from the unstable radiation dominated epoch, $f_2$, continues toward unstable matter dominated epoch, $f_4$, and ends at the stable dark energy dominated point, $f_6$. In Fig. \ref{Range}, left panel,  we plotted the behaviour of density parameters with respect to the cosmic time for an appropriate set of initial conditions. We set initial conditions deep in radiation dominated era in a way that universe evolves through standard cosmological epochs and reaches the present value of $\Omega_{\text{m0}} \approx 0.05$ and $\Omega_{\text{r0}} \approx 10^{-5}$. Note that since the theory is designed to address the dark matter problem, we only include the baryonic part of matter for $\Omega_{\text{m0}}$. In Fig. \ref{Range}, right panel, we demonstrate the behaviour of $\omega_{\text{eff}}$ and $\omega_{\text{DE}}$ for the same initial conditions that used for density parameters. It should be noted that the singularity in the dark energy equation of state is not a matter of problem since $\omega_{\text{eff}}$ that determines the physical behaviour of the cosmic observable quantities, smoothly evolves with the cosmic time.

In the following we study two different limits in MOG. First we assume that the vector field does not contribute to the evolution of the cosmic fluid. In other words, we suppose that its effects appear only in the local gravitating systems like galaxies. In fact, in cosmology, baryonic matter is treated as dust, and the averaging scale of homogeneity and isotropy is about $\sim 100 \text{Mpc}$. On the other hand, it is shown in  \cite{Moffat:2013sja} and \cite{Moffat:2013uaa} that the current magnitude of the Compton wavelength of the vector field is about $ \mu^{-1}\sim 24\, \text{kpc}$. It is important to mention that, in fact, the $\mu$ scalar field plays the role of the Compton wavelength for the vector field. Consequently this wave length is not constant and changes with time. Furthermore it depends on the self-gravitating system and one cannot find a universal value for it \cite{haghi}. However for our purpose here we use $ \mu^{-1}\sim 24\, \text{kpc}$ as a crude estimation.

Therefore one may expect that the influence of $\phi_0$ is confined entirely inside self-gravitating systems constructed from dust particles (i.e. galaxies).  So from a cosmological perspective $\phi_0$  plays no role and the value of its average $\langle\phi_0\rangle$ is irrelevant. From this point of view, it is appropriate to set $\phi_0=0$ when dealing with the cosmological behaviour \cite{n2}. In this case the theory reduces to a scalar-tensor theory of gravity with two scalar fields. 

In another limit, we set to zero the constant $\Lambda$ and investigate two cases, i.e. $\Lambda=0$, $\phi_0=0$  and $\Lambda=0$, $\phi_0\neq 0$. In this case, we are interested to check the potential of MOG to work as a dark energy model rather than following the original motivation of an alternative to the dark matter problem. More specifically, we investigate the possibility if MOG could explain the cosmic speed up without an explicit cosmological constant.

\section{MOG without the Proca vector field}\label{sec-4.1}
In order to check the behaviour of MOG in the absence of the vector field, we set $\phi_0$ to zero. In this case the dimension of the phase space reduces by one. Also it is easy to show that $Q$ is zero and one can write the dynamical variable $m$ in terms of the rest variables as $m^2=1-y-r-x^2-z+\frac{4\pi}{3}z^2$. It turns out that there are three fixed points $(x,y,r,z)$,
\begin{equation}
\begin{split}
&\!\!\!\!\!p_{1}: (0,0,1,0)\qquad  ~~~~~~~~  \omega_{\text{eff}}=\frac13\\&
\!\!\!\!\!p_{2}:(0,0.96,0,0.04)\qquad   \omega_{\text{eff}}=-0.01 \\&
\!\!\!\!\!p_{3}:(0.98,0,0,0.04) \qquad  \omega_{\text{eff}}=-1.00
\end{split}
\end{equation}
Let us discuss these points in turn.
\begin{itemize}
	\item {\textbf{\textit{ $p_1$: Radiation dominated era:}}}
\end{itemize}
This fixed point corresponds to a standard radiation dominated epoch for which $a(t) \propto t^{0.50}$. The eigenvalues in this case are $(-2,-1,2,1)$ that shows $p_1$ is unstable as expected. In this era both scalar fields  $G(t)$ and $\mu(t)$ remain constant with time. 
\begin{itemize}
	\item {\textbf{\textit{ $p_2$: $G$-Matter dominated era:}}}
\end{itemize}
This is an unstable matter dominated epoch since the eigenvalues of the stability matrix are $(-2.96,1.48,-1.48,-1)$. The existence of just one positive eigenvalue implies  instability of the fixed point. In this era, $\omega_{\text{eff}}=-0.01$, which results in $a(t) \propto t^{0.68}$. Note that $p_2$ behaves exactly as the point $f_4$ in previous section and both are close to the standard case for which $a(t) \propto t^{0.67}$. The running of the scalar field $G(t)$ starts from the matter dominated epoch and in fact, in this era, we have $G(t) \propto t^{0.03}$ and $\mu(t)$ is approximately constant.

 \begin{itemize}
	\item {\textbf{\textit{ $p_3$: $\Lambda G $  era:}}}
\end{itemize}
This point corresponds to a $ \Lambda G$ epoch for which the eigenvalues are $(-5.92,-3.96,-2.96,-2.96)$. The negative eigenvalues of the stability matrix show that this is a stable epoch. In this case $\omega_{\text{eff}}= -1.00$ (up to two digits, as mentioned before) as one expects for a viable late time solution. Note that the running of the field $G(t)$ that started in the matter dominated epoch, continues until late time and evolves as  $e^{0.04 t}$. On the other hand $\mu(t)$ is still constant. 

It is important to mention that in the absence of the vector field, MOG still possesses  viable cosmic epochs. Surprisingly, the fixed points $p_{1},p_2$ and $p_3$ are the same as $f_2$, $f_{4}$ and $f_{6}$, respectively. In other words, as we already mentioned, the vector field does not significantly change the cosmic evolution. However it is necessary to mention that in the presence of the vector field we have two fixed points for each epoch while in its absence there is only one fixed point for each phase.

\section{MOG as a dark energy model }\label{sec-4.2}
\begin{table*}[!]\renewcommand{\arraystretch}{1.5}
\small
\begin{center}
\begin{tabular}{|l c c c c|}
\hline
Point& $(y,r,z,m) $ & ~~~~$(\Omega_{\text{m}},\Omega_{\text{r}},\Omega_{\text{G}},\Omega_{\mu}) $~~~~ & $\omega_{\text{eff}}$ & Stability\\ \hline \hline 
$b_1$ & $(0,1,0,0)$ & (~0,~1,~0,~0)&$\frac13$& unstable\\ 
$b_2$ & $(0,1.60,-0.53,-2.05)$ & (~0,~1.60,~-1.72,~-4.19)&0.51 & unstable\\ 
$b_3$ & $(0.96,0,0.04,-6.36)$ & (~0.96,~0,~0.03,~0)&-0.01 & stable\\ 
$b_4$ & $(1.71,0,-0.74,-3.07)$ & (~1.71,~0,~-3.03,~-9.42)&0.25 & unstable\\ 
\hline
\end{tabular}
\caption{Fixed points and their stability for $(\phi_0 \neq 0, \Lambda =0)$}\label{tab2}
\end{center}
\end{table*}
This section is devoted to answering the question whether MOG can behave as a dark energy model. In fact, as we already mentioned, in this theory there are two scalar fields, $G(t)$ and $\mu(t)$ , and one vector field, $\phi_{\alpha}(t)$ , besides the metric tensor. So, the logical question is whether the extra fields can play the role of dark energy, as they do for the dark matter. In order to check this possibility, we remove the cosmological constant, $\Lambda$, and investigate the governing equations \eqref{zprime} when $x=0$. Note that removing $x$ reduces the dimension of the phase space by one.  Furthermore, since the theory shows a standard sequence of cosmological epochs, regardless of the presence of the vector field, we investigate the viability of MOG as a dark energy model in both cases, i.e. when $\phi_0 \neq 0$ and $\phi_0 =0$.

In the case of the vanishing vector field, the phase space is three dimensional $(y,r,z)$. The fixed points are:
\begin{equation*}
\begin{split}
&\!\!\!\!\!d_{1}: (0,1,0)\qquad  ~~~~~~~~  \omega_{\text{eff}}=\frac13\\&
\!\!\!\!\!d_{2}:(0.96,0,0.04)\qquad   \omega_{\text{eff}}=0.01
\end{split}
\end{equation*}

Now, we have only two fixed points. $d_1$ is a radiation dominated epoch for which eigenvalues are $(-2,-1,1)$, i.e. an unstable radiation dominated era. On the other hand $d_2$ is a stable matter dominated era with eigenvalues $(-3, -1.5, -1)$. There is no stable late time fixed point. Therefore, because of the stability of the matter dominated phase and the absence of a stable de Sitter universe, we conclude that MOG without cosmological constant and without the  vector field does not lead to a viable cosmological model.

Now let us bring back the vector field and re-analyse the system in the absence of $\Lambda$. The results are shown in Table \ref{tab2}. In this case, we have four fixed points $b_1$-$b_4$ and the corresponding eigenvalues are $(-2., -1., -1., 1.)$, $(-2.53,-1.27,1.27,1)$, $(-3, -1.48, -1.48, -1)$ and $(-3.89,2.02,-1.87,-1.)$ respectively. The stability of the matter dominated phase $b_3$ is not a big problem since there is another unstable matter dominated epoch $b_4$. There is also a standard radiation dominated era $b_1$, and an unusual radiation dominated era $b_2$. The main problem here is the absence of a standard late time fixed point. Moreover, the unstable matter dominated point $b_4$ differs drastically from the standard case. Therefore, one may simply conclude that without $\Lambda$, MOG does not work as a valid cosmological model.

To summarize this section, we reiterate that if we set $\Lambda=0$ in the standard version of MOG, in which the self-interaction potential of the scalar fields are zero, then MOG does not work properly at the cosmological level. In other words, MOG cannot play the role of dark energy, although one may construct a model of MOG including non-zero self-interacting potentials to explain the cosmic speed-up without cosmological constant. This issue needs further investigation. Another modification can be achieved by introducing new constants as coefficients for the kinetic terms in the scalar fields Lagrangian. This case is straightforward and can be considered as an attempt to construct a dark energy version of MOG. Our modified MOG, as we will see,  introduces a new constant $c$ for $G$ that plays the same role as the Brans-Dicke constant $\omega$.
 Therefore before moving on to discuss some observational constraints on the standard version of MOG, we study the above mentioned extended version of MOG.

For completeness we keep the cosmological constant. In fact it turns out that, similar to what we discussed in sections \ref{sec-2} and \ref{sec-4.1}, to find the fixed points when $\Lambda=0$ it is enough to set $x=0$ in the main fixed points which have been obtained for non-zero cosmological constant case. Keeping the cosmological constant in action \eqref{action2} enables us to investigate another question. In fact, we saw that in the standard version of MOG, the matter dominated point $f_4$ is slightly different from the standard matter dominated case and has a negative effective equation of state parameter. The  extension we propose now may help to reconstruct a standard matter phase.  Moreover, in the next section, we will compute the angular size of the sound horizon and use this modification in order to make a more consistent model.  Keeping in mind our purposes, let us start with the following modified action, which, as mentioned before, we call it mMOG throughout the paper
\begin{equation} \label{action2}
\!\!S\!\!=\!\!\int \!\! \sqrt{-g} d^{4}x \bigg[
 \frac{(R\!+\!2\Lambda)}{16 \pi G}  \!\!+\!\!\frac{\omega_{0}}{4 \pi}\left(\frac{1}{4} B_{\mu \nu}B^{\mu \nu} \!\!+\!\!V_{\phi}\right)
 \!\!+\!\!\frac{1}{2 G}\left( \frac{c}{8\pi}\frac{\nabla_{\mu} G \nabla^{\mu} G}{G^2}\!+\!\frac{c'\nabla_{\nu} \mu \nabla^{\nu} \mu}{\mu^2}\right)\bigg] \!\!+\!\!S_{\text{M}}
\end{equation}

       \begin{table}[!] \renewcommand{\arraystretch}{1.5}\renewcommand{\tabcolsep}{0.01cm}
       \scriptsize
       \begin{center}
         \begin{tabular}{|cccc|}
         \hline   
               Point &  $\textbf{\large{}(}\Omega_{\Lambda},\Omega_{\text{m}},\Omega_{\text{r}}, \Omega_{\text{G}} ,\Omega_{\mu} \textbf{\large{})}$  &  $\omega_{\text{eff}}$  &  Stability  \\
         \hline
         \hline
         $F_1$                &  $(0,0,\frac{(2 c+4 \pi -3)^2}{(3-2 c)^2},\frac{\pi  (72-52 c)}{3 (3-2 c)^2}, \frac{-4 \pi }{3})$       &  $\frac{8 \pi }{6 c-9}+\frac{1}{3}$  &  unstable         \\
       
        $ F_2$                    &  $(0,0 ,1 , 0,0)$     &  $\frac13$  &  unstable  \\
         $F_3$                   &  $(0, \frac{6 c^2+(30 \pi -17) c+6 \left(2-7 \pi +6 \pi ^2\right)}{6 (c-1)^2},0,-\frac{(6 \pi -1) ((54 \pi -1) c-48 \pi )}{48 \pi  (c-1)^2},-3 \pi )$     & \!\!  $\frac{1-6 \pi }{3-3 c}$ \!\! & \!\!\! unstable for $c \!\! \neq \!\! 1$ \!\!\!\! and \!\!\!\! $c \!\! \neq \!\! \frac{(4 - 6 \pi)}{3}$  \\
         $F_4$          &  $(0,\frac{6 c^2-17 c+12}{6 (c-1)^2},0,\frac{1}{c-1}-\frac{c}{48 \pi  (c-1)^2},0)$ \!\! & \!\! $\frac{1}{3-3 c}$ \!\! & \!\! unstable if $c \neq 1$ \!\! and \!\! $c \neq \frac43$        \\
       $ F_5$  & \!\! $(\frac{(3 c+6 \pi -4) (2 c+8 \pi -3)}{6 (-c+2 \pi +1)^2},0,0,\frac{(12 \pi -1) (-60 \pi  c+c+48 \pi  (1+2 \pi ))}{48 \pi  (c-2 \pi -1)^2},-\frac{\pi  (7-6 c)^2}{3 (c-2 \pi -1)^2})$  &  $-1$   &  unstable   \\
        $ F_6$         &  $(\frac{c^2-\frac{17 c}{6}+2}{(c-1)^2},0,0,\frac{1}{c-1}-\frac{c}{48 \pi  (c-1)^2},0)$  &  $-1$ &stable for $c<1$ or $c>\frac32$   \\
         \hline
         \end{tabular}
         \caption{Fixed points and their stability for generalized MOG}\label{table3}
         \end{center}
       \end{table}

In fact it turns out that one can include only two independent dimensionless coefficients, $c$ and $c'$. Any constant in front of $R$ can be absorbed by a redefinition of the $G$ field. As already mentioned,  $\omega_0$ can be eliminated by a redefinition of $\mu$. The action \eqref{action2} recovers the standard action by setting $c$ and $c'$ to $8\pi$ and $1$, respectively.  

 As we saw in the previous sections, the $\mu$ field does not play a key role in the cosmic expansion. In other words, as it is clear in Fig. \ref{Range} (left panel) the $\mu$ contribution,  $\Omega_{\mu}$, to the total cosmic energy budget is negligible and remains constant compared with the other components. It is also emphasized in \cite{Moffat:2014bfa} that $G$ plays a more important role in the cosmology of MOG than the $\mu$ field. On the other hand, the weak field limit of theory shows that $\mu$ is the field that specifies the length correspond to the mass scale of vector field while the field that plays the main role for increasing the gravitational potential is $G$ \cite{Moffat:2013uaa}. In fact, a straightforward consideration of the gravitational force between point masses in the weak field limit of MOG also reveals the importance of the field $G$, compared to $\mu$, at large distances. Finally, a  constant coefficient for the kinetic term of $\mu$ can be absorbed in the definition of $\Omega_{\mu}$ and will not significantly influence the calculations of the angular size of the sound horizon. That is why we set the coefficient $c^{\prime}$ to $1$ for the rest of the calculations and focus solely on $c$.

It is worth mentioning  that the current value of Newton's constant $G_{\text{N}}$, is not affected by the dimensionless coefficient $c$. In fact, a numerical analysis of dynamical variable $z=\frac{{G^\prime}}{G}$, using the field equations, reveals that $\frac{G_0}{G(t)}$ is the quantity that is specified, where $G_0$ is an integration constant and its value can be matched with the value of Newton's constant $G_{\text{N}}$.

Although we proceed the study irrespective of the sign of $c$, for the reason discussed in Appendix \ref{app} we are interested in behavior of the theory for positive $c$. On the other hand, it is necessary to be sure that the existence of $c$ does not affect the weak field limit of the theory. To check this point, using the method in \cite{ra}, we showed that $c$ will be absorbed in the definition of the $\alpha$ parameter which is related to $G$. Therefore the existence of $c$ does not disturb the main feature of theory in the weak field limit, which is the increase of  the strength of gravity.

  The generalized field equations are:
 \begin{equation}
\begin{split}
& H^2 =\frac{8 \pi G}{3} \rho + \frac{\Lambda}{3} + \frac{\dot{G}\dot{a}}{G a} 
 - \frac{4 \pi}{3}( \frac{\dot{\mu}^2}{\mu^2} + \frac{c}{8 \pi} \frac{\dot{G}^2}{G^2} )+ \frac{1}{3} G  \mu^2 \phi_0^2\\&
\frac{\ddot{a}}{a}=-\frac{4\pi G}{3}(\rho+3 p)+\frac{1}{2}\frac{\ddot{G}}{G}+(\frac{c -3}{3})\frac{\dot{G}^2}{G^2}-\frac{2 G \mu^2 \phi_0^2 }{3}
+\frac{\Lambda}{3}+\frac{1}{2}\frac{\dot{G}}{G}H+\frac{8 \pi}{3}\frac{\dot{\mu}^2}{\mu^2}\\&
\frac{\ddot{G}}{G}= \frac{3}{c}\left(\frac{\ddot{a}}{a}+ H^2 \right)-3 H \frac{\dot{G}}{G} +\frac{3\dot{G}^2}{2 G^2} -\frac{4 \pi \dot{\mu}^2}{c \mu^2}-\frac{\Lambda}{c}
\end{split}\label{gfe1}
\end{equation}

The field equations of $\mu$ and the vector field and other definitions are the same as in section \ref{sec-1}. The only difference is that $c$ appears in the definition of $\Omega_{\text{G}}$ as $ \Omega_{\text{G}}=z-c \frac{z^2}{6} $. Therefore, it is straightforward to find the critical points of the relevant dynamical system. For simplicity, in Table \ref{table3} we have listed the density parameters $(\Omega_{\Lambda},\Omega_{\text{m}},\Omega_{\text{r}}, \Omega_{\text{G}} ,\Omega_{\mu})$ of the critical points instead of the fixed points $(y,r,x,z,m)$.
Let us now briefly discuss  these fixed points, specially  the matter dominated era. $F_1$ and $F_2$ are unstable radiation dominated eras with eigenvalues $(1,\frac{4 \pi  (8 \text{c}+1)-6}{16 \pi  \text{c}-3},\frac{6-8 \pi  (4 \text{c}+1)}{16 \pi  \text{c}-3},\frac{3-4 \pi  (4 \text{c}+1)}{16 \pi  \text{c}-3},\frac{4 \pi  (4 \text{c}+1)-3}{16 \pi  \text{c}-3})$ and $(0,1,0,0,0)$ respectively. However it is clear that $F_2$ is a standard radiation dominated point. In fact  MOG possesses a valid  radiation dominated epoch $F_2$ irrespective of the value of $c$. On the other hand $F_3$ is a matter dominated point in which the scalar fields $\mu$ and $G$ play a role. Considering the eigenvalues of the stability matrix (which are too long to be written here) it turns out that this epoch can be unstable if $c\neq 1$ and $c\neq \frac{(4 - 6 \pi)}{3}$. We reiterate that in the  standard version of MOG, $c=8 \pi$ and therefore $F_3$ is unstable and $a(t) \propto t^{0.5}$, which is drastically different from the standard case. We are not going to discuss more about this fixed point, since MOG possesses another fixed point, $F_4$, which is very similar to a standard matter dominated epoch.

$F_4$ is a matter dominated point where the scalar field $G$ is also important. In this phase, the scale factor grows as $a(t) \propto t^{\frac{2 (c-1)}{3 c-4}}$ . The eigenvalues of the stability matrix for $F_4$ are $(-3,-1,\frac{4-3 c}{2 (c-1)},\frac{4-3 c}{2 (c-1)},\frac{4-3 c}{2-2 c})$. Therefore this point is unstable provided that $c\neq\frac{4}{3}$ and $\text{c}\neq1$. Although for $\frac{c}{8 \pi}= 1$ this point is reminiscent of the standard matter dominate phase, as we showed in the previous section, it does not behave exactly as standard model, i.e. $t^{2/3}$. However it seems that one gets an almost standard matter dominated epoch by choosing large values of $c$.

Furthermore, $F_5$ can be considered as a phase in which $\Lambda$ plays an important role. However, this critical point cannot be considered as a stable late time solution since all the eigenvalues (too long to be written here) are not negative. On the other hand, $F_6$ is an acceptable late time solution. In fact eigenvalues for $F_6$ are $(\frac{4-24 \pi  \text{c}}{8 \pi  \text{c}-1},\frac{4-24 \pi  \text{c}}{8 \pi  \text{c}-1},\frac{4-24 \pi  \text{c}}{8 \pi  \text{c}-1},\frac{7-48 \pi  \text{c}}{8 \pi  \text{c}-1},\frac{5-32 \pi  \text{c}}{8 \pi  \text{c}-1})$. Therefore $F_6$ is stable if $c<1$ or $c >\frac{3}{2}$. For this epoch, irrespective of the value of $c$, we have $\omega_{\text{eff}}=-1$. In other words, the coefficient of the kinetic energy of the $G$ field does not change the expansion rate of the cosmos at  late times.

Therefore, from the stability point of view, one can increase the value of $c$ to suppress the deviations from the standard matter dominated phase without disturbing the existence of other viable cosmic eras. On the other hand, it is necessary to mention that introducing the coefficient $c$ does not help to find an accelerated expansion without cosmological constant. In other words, as it is clear from the fixed points in Table \ref{table3}, no value of $c$ does  lead to a new fixed point in which $\Lambda=0$ and $\omega_{\text{eff}}=-1$. Consequently this simple attempt to make a dark energy version of MOG without a cosmological constant fails, although helps to construct a more viable matter dominated phase.

\section{Angular size of the sound horizon in MOG}\label{sec-3.1}
\begin{figure}
\centerline{\includegraphics[width=10cm]{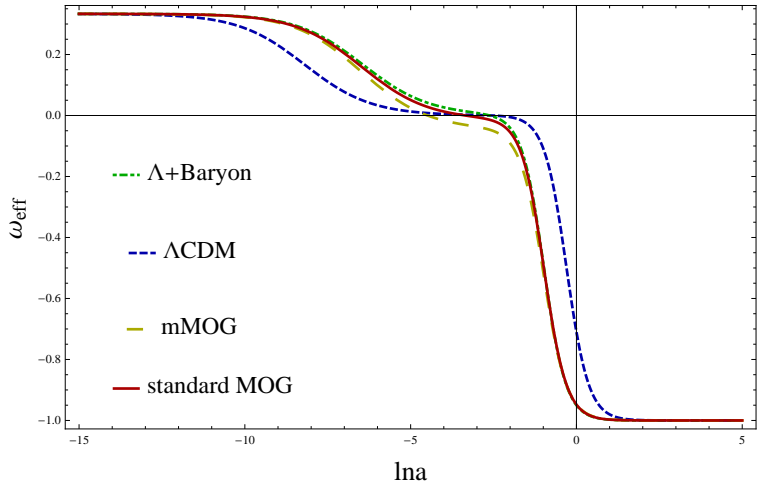}}
\caption[]{The evolution of $\omega_{\text{eff}}$ for four models, $\Lambda$CDM (the blue dashed line), $\Lambda$ + baryons (the green dot-dashed line), standard MOG (solid line) and mMOG ($c=0.33 \times 8 \pi$) (long-dashed line). While standard MOG and $\Lambda$ + baryons are very similar, one can see the deviation between mMOG and other three models, in the matter dominated era, due to having negative equation of state in this phase. In the case of standard MOG, mMOG and $\Lambda$+Baryon we used the initial condition that result in $\Omega_{\text{m0}} \approx  0.05$, while for $\Lambda$CDM we set initial conditions at present as $\Omega_{\text{m0}} \approx 0.3$.  In all cases we set $\Omega_{\text{r0}} \approx 10^{-8}$.}
\label{eff}
\end{figure}

We found in section \ref{sec-2} that the standard version of MOG possesses a viable sequence of cosmological eras, although the matter dominated era is not exactly standard. More specifically, we found $\omega_{\text{eff}}=-0.01$ and $a(t) \propto t^{0.67}$ for the matter dominated era. The negative sign for the $\omega_{\text{eff}}$ in matter dominated era shows the effect of the existence of extra fields in this era. In fact, this solution is reminiscent of the $\phi$MDE era in the $f(R)$ theories \cite{Amendola:2006kh}. Since the theory has an acceptable sequence of cosmological epochs, one can go further and ask about other cosmological constraints. For this purpose, we compute the angular size of the sound horizon, $\theta_{\text{s}}$, in MOG
\begin{equation}
 \theta_{\text{s}} = \frac{\int_{z_{\text{dec}}}^{\infty} \frac{c_{\text{s}}(z) dz}{H(z)}}{\int_0^{z_{\text{dec}}}\frac{dz}{H(z)}}
 \end{equation}
 Where $c_s^2(z)=1/[3(1+\frac{3 \rho_{\text{b}}}{4 \rho_{\gamma}})]$ it is the adiabatic baryon-photon sound speed and $z_{\text{dec}}$ is the redshift at decoupling. The current observational value of this $\theta_{s}$ measured by Planck 2015 is $100 ~ \theta_{\text{s}}= 1.04105 \pm 0.00046$ radians \cite{Riess:2004}. This value is obtained assuming a constant equation of state parameter. As it is clear in Fig. \ref{eff}, $\omega_{\text{eff}}$ is constant in radiation dominated universe for a wide time interval. Therefore as in \cite{Amendola:2006kh}, we use this observational value in MOG. It is also important to mention that as the standard conservation law for $\rho_{\text{r}}$ and $\rho_{\text{m}}$ is satisfied in MOG, $z_{\text{dec}}$ is unchanged and we use the same value as in $\Lambda$CDM in our calculations.

\begin{figure}
\centerline{\includegraphics[width=10cm]{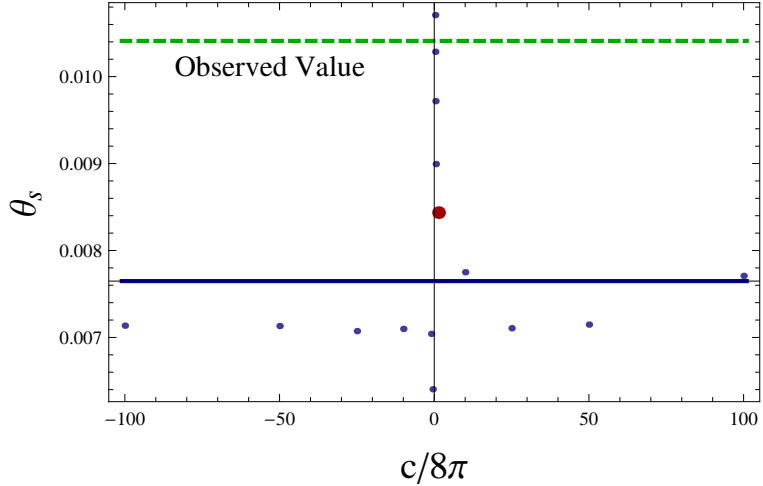}}
\caption[]{The angular size of the sound horizon for different values of $\frac{c}{8 \pi}$. The dashed curve shows the observed value. The solid line  corresponds to a pure-GR toy model without  dark matter component. Dots represent  different values of $c$ in mMOG.  The large red dot illustrates the value of $c$ equal to $8 \pi$, which is equivalent to the standard version of MOG.}
\label{cplot}
\end{figure}

Now, let us calculate $\theta_{\text{s}}$ for the standard version of MOG which is introduced in section \ref{sec-1}. To do so, we set the initial conditions in a way that we find observed values for baryonic matter and radiation at the present, i.e. $\Omega_{\text{m0}} \approx 0.05$ and $\Omega_{\text{r0}} \approx 8 \times 10^{-5} $. Note that since MOG is going to ignore the presence of dark matter, the present value for matter contains only the baryonic part of the matter content of the Universe. Our calculations show that for an interval of initial conditions that leads to $0.04<\Omega_{\text{m0}}<0.05$ and $7 \times 10^{-5} <\Omega_{\text{r0}}< 9 \times 10^{-5}$, the angular size of the sound horizon lies in the interval $0.0080  \text{rad} <\theta_{\text{s}}< 0.0082 \text{rad} $. We have shown the maximum value of $\theta_{\text{s}}$ with a large red point in Fig. \ref{cplot}. As it is clear from the figure, this range is not consistent with the above mentioned observational values of $\theta_{\text{s}}$. More specifically $\theta_{\text{s}}$ in MOG is 19 \% smaller than the observed value and more than $100 \sigma$ error away from the Planck data. Therefore, in the following we use the modified version of MOG, dubbed mMOG, introduced in section \ref{sec-4.2} to see if the new constant $c$ in the action \eqref{action2} can help improve the agreement with Planck data. In the previous section we showed that by increasing this parameter, one may recover an exact matter dominate era in MOG. In what follows, we study the effect of this parameter on the value of $\theta_{\text{s}}$.
The results are again in Fig. \ref{cplot}. The dashed line shows the observed value of $\theta_{\text{s}}$ in $\Lambda$CDM, with error bars smaller than the line thickness. The solid line belongs to a toy model in GR with $\Lambda$ without dark matter component. We set the current value of the matter density equal to baryonic part, i.e. $0.05$, and calculate $\theta_{\text{s}}$. We will compare MOG with this toy model in order to gain more intuition about the cosmological behaviour of MOG.

The dots in Fig. \ref{cplot} correspond to $\theta_{\text{s}}$  for mMOG with different values for $\frac{c}{8 \pi}$. For every $c$ we set initial conditions at $\ln a \approx -20$ in order to find the observed value for $\Omega_{\text{m}}$ and $\Omega_{\text{r}}$ at present. As the plot suggests, mMOG cannot fit the data for large values of $c$. In fact, for $c/8 \pi > 50$, the deviation of $\theta_{\text{s}}$ from the observed value becomes larger and tends to a same magnitude as in $\Lambda$ + Baryon toy model, i.e. $\theta_{\text{s}}\simeq 0.0076$ radians. Now, let us discuss about smaller values of $c$. Choosing $c = 8 \pi \times 0.33$ results in   $\theta_{\text{s}}=0.0103$, which is very close to the observed value as shown in Fig. \ref{cplot}. It is important to clarify that for $c=0.33 \times 8 \pi$, using Table \ref{table3} for the critical point $F_4$, one finds $\omega_{\text{eff}}=-0.05$ that results in $a(t) \propto t^{0.70}$. As Fig. \ref{cplot} suggests, standard MOG behaves cosmologically like $\Lambda$CDM without dark matter, instead of an alternative theory for dark energy. It also means that the extra fields of MOG cannot reproduce the role of dark matter in the sense that the theory behaves like $\Lambda$CDM without dark matter. This fact can also be seen from the time evolution of $\omega_{\text{eff}}$. In Fig. \ref{eff}, we illustrate the evolution of $\omega_{eff}$ for 4 models, i.e. $\Lambda$CDM, $\Lambda$+ Baryons, standard MOG and mMOG. It is evident from Fig. \ref{eff} that standard MOG and  $\Lambda$ + baryons are very similar. However it is clear that there are significant differences between $\omega_{\text{eff}}$ in MOG and $\Lambda$CDM. More specifically as Fig. \ref{Range} , left panel, suggests, matter dominated phase occurs approximately in the interval $-5<\text{ln} a<-2$. In this interval in Fig. \ref{eff}, the evolution of $\omega_{eff}$ for mMOG is different from that of the other three models.  In fact, since mMOG has a negative $\omega_{\text{eff}}$ during the matter dominated era, one expects to see deviation from other models.

It is also instructive to compare the age of the universe for these models. One may simply integrate equation \eqref{lan} to find the age. It should be noted that the dynamical system approach has already provided us with $\omega_{\text{eff}}$ numerically. The slower expansion rate in the matter dominated phase of mMOG compared with the standard case, may substantially increase the age. As expected, our calculations reveals the fact that mMOG predicts an older universe than $\Lambda$CDM. While the age of universe in $\Lambda$CDM is around $14$ billion years, mMOG with $c=0.33 \times 8 \pi$ results in a $22$ billion years old universe. On the other hand for standard MOG and $\Lambda$ + Baryons the age is around 21 billion years.

\section{Conclusion}\label{sec-5}
We investigated the cosmological behaviour of MOG using the phase space analysis method. This approach provides a fast method to solve the complicated field equations numerically and we used the solutions to calculate the angular size of the sound horizon.  We found that MOG, in its standard form, possesses acceptable cosmological epochs. In fact, the Universe starts from a standard radiation dominated era $f_2$, continues toward an unstable matter dominated epoch $f_4$ and finally ends in the late time attractor $f_6$. However the matter dominated point $f_4$ is slightly different from the standard case. In section \ref{sec-4.1}, we turned off the effects of the vector field and showed that MOG still has a viable sequence of cosmological eras in the absence of the Proca vector field. In section \ref{sec-4.2} we checked the viability of MOG as a dark energy model by setting $\Lambda=0$ in the field equations. We showed that the extra fields of MOG cannot play the role of dark energy.
In section \ref{sec-4.2}, we introduced a modified version of MOG, called mMOG, by adding new constants as coefficients for the kinetic terms of the scalar fields. Although mMOG cannot accelerate the expansion in  absence of $\Lambda$, we find that by increasing the value of the parameter $c$  the cosmic scale factor during the matter dominated era gets close to the standard matter dominated era, i.e. $a(t)\propto t^{2/3}$. 

The fact that standard MOG possesses a valid sequence of standard cosmological epochs prompted us to compare MOG with a very precise cosmological observation, i.e. the angular size of the sound horizon $\theta_{\text{s}}$. Our calculations show that $\theta_{\text{s}}$ in the standard MOG is 19 \% smaller than the observed value, way off the experimental errors. However, for mMOG, with a new constant $c$, we found that  $c=0.33\times 8\pi$ leads to an appropriate value for $\theta_{\text{s}}$ consistent with observations. On the other hand, in this case, the  matter dominated phase becomes slightly different from the standard matter dominated phase, which could cause deviations in the linear perturbation growth.

In conclusion, this paper shows that, with or without an explicit cosmological constant, standard MOG is not a cosmologically acceptable model. A slightly modified and tuned version, called mMOG, fits the observed value of the sound horizon but contains a matter dominated epoch for which $a(t) \propto t^{0.70}$, that might have consequences on linear perturbation growth.

\acknowledgments
This work is supported by Ferdowsi University of Mashhad under Grant NO. 39640(04/11/1394). We would like to thank the anonymous referee for useful comments. SJ thanks the Institute for Theoretical Physics, University of Heidelberg where part of this work was carried out. LA acknowledges support from the DFG project TR33 "The Dark Universe". MR would like to thank Ahmad Ghodsi for insightful discussions.

\appendix
\section{Appendix}\label{app}
\renewcommand{\thesection}{\Alph{section}}
\numberwithin{equation}{section}
In this appendix, we study the classical stability of MOG and show that there is no tachyonic instability in the theory. As we mentioned in Sec. \ref{sec-1}, the scalar fields $G$ and $\mu$ appear with wrong sign for the kinetic terms. This means that  the theory cannot be taken to be a fundamental quantum field theory.  

In order to discuss the stability of the theory, it is convenient to define a scalar field $\Phi$ which is related to $G$ as $G=\frac{1}{\Phi}$. Furthermore, in what follows, for simplicity, we ignore $S_{\text{M}}$.

With the above mentioned redefinition and assumption, the action (\ref{action}) takes the following form
 \begin{equation}
 S=\int \sqrt{-g} d^4 x \Big[ \frac{\Phi R}{16 \pi} +\frac{\Lambda \
   \Phi}{8 \pi}+ \frac{\nabla_{\alpha}\Phi \nabla^{\alpha}\Phi}{2 \
 \Phi} + \frac{\Phi \nabla_{\alpha}\mu \nabla^{\alpha}\mu}{2 \mu^2}  +\frac{B_{\alpha \delta} B^{\alpha \delta}}{16 \pi}  -  \frac{\mu^2 \phi_{\alpha} \
  \phi^{\alpha}}{8 \pi}\Big]
 \label{reaction}
 \end{equation} 
By varying this action with respect to the metric tensor, one  finds the following field equation

 \begin{equation}
 \begin{split}
 & \frac{\Phi \
  R_{\alpha \beta}}{16 \pi}  -  \bigr(\frac{R}{32 \pi} + \frac{\Lambda}{16 \pi} \bigr) \Phi g_{\alpha \beta}  -  \frac{\mu^2 \phi_{\alpha} \phi_{\beta}}{8 \pi} + \
\frac{g_{\alpha \beta} \mu^2 \phi_{\gamma} \phi^{\gamma}}{16 \pi}  +\frac{B_{\alpha}{}^{\gamma} B_{\beta \gamma}}{8 \pi} -  \
 \frac{B_{\gamma \delta} B^{\gamma \delta} g_{\alpha \beta}}{32 \pi} \
      \\&
+ \frac{\nabla_{\alpha}\Phi \nabla_{\beta}\Phi}{2 \Phi} + \frac{\Phi \nabla_{\alpha}\mu \nabla_{\beta}\mu}{2 \mu^2} -  \frac{\nabla_{\beta}\nabla_{\alpha}\Phi}{16 \pi} + \bigr( \frac{ \Box\Phi}{16 \pi} -  \frac{ \nabla_{\gamma}\Phi \nabla^{\gamma}\Phi}{4 \Phi} -  \frac{\Phi \nabla_{\gamma}\mu \nabla^{\gamma}\mu}{4 \mu^2}\bigr)g_{\alpha \beta}=0
 \end{split}
 \label{eom}
 \end{equation}
where $\Box=\nabla_{\gamma}\nabla^{\gamma}$. Note that, in a homogeneous and isotropic background, the anti-symmetric tensor $B_{\alpha \beta}$ vanishes. Using the trace of \ref{eom}, one can find $R$ as 
 \begin{equation}
 R=  -4 \Lambda + \frac{2 \mu^2 \phi_{\alpha} \phi^{\alpha}}{\Phi} + \frac{3 \
   \Box\Phi}{\Phi} -  \frac{8 \pi \nabla_{\alpha}\Phi \nabla^{\alpha}\Phi}{\Phi^2} -  \frac{8 \pi \nabla_{\alpha}\mu \nabla^{\alpha}\mu}{\mu^2}
   \label{R}
   \end{equation}
On the other hand the variation with respect to $\Phi$ gives 
\begin{equation}
\frac{\Box\Phi}{\Phi}= \frac{\Lambda}{8 \pi} + \frac{R}{16 \pi} + \frac{\nabla_{\alpha}\Phi \nabla^{\alpha}\Phi}{2\Phi^2} + \frac{\nabla_{\alpha}\mu \nabla^{\alpha}\mu}{2 \mu^2}
\label{Fi}
\end{equation} 
substituting $R$ using (\ref{R}), we can rewrite (\ref{Fi}) as
 \begin{equation}
\Box\Phi +\frac{2 \Lambda }{ 16 \pi -3 }\Phi + \frac{2 \mu^2 \phi_{\alpha} \phi^{\alpha}}{3 - 16 \pi} =0.
 \label{Fir}
 \end{equation}
Keeping in mind the metric signature used in this paper, equation \eqref{Fir} is a Klein-Gordon like equation, in which the term $\frac{2 \Lambda }{ 16 \pi -3 }$ appears as a positive mass. This explicitly shows that $\Phi$, or equivalently $G$, is not a tachyon. In the modified version of MOG, mMOG, the mass term is $\frac{2 \Lambda }{16 \pi c -1}$, and in order to have positive mass, one should keep the constraint $c > \frac{1}{16 \pi}$.  Let us also find the field equation of $\mu$. Variation with respect to $\mu$ yields
\begin{equation}
\Box\mu + \frac{\phi_{\alpha} \phi^{\alpha}}{4 \pi \Phi} \mu^3 +  \frac{\nabla_{\alpha}\mu \nabla^{\alpha}\Phi}{\Phi} - \frac{\nabla_{\alpha}\mu \nabla^{\alpha}\mu}{\mu}=0
\label{mu}
\end{equation}
The $\mu^3$ term can be considered as a derivative of an effective potential, which then would be a power law potential of type $\mu^4$, with the positive sign. Therefore one may expect that this potential is also stable, and consequently there is no tachyonic instability for $\mu$.

In the following, for the sake of completeness, we briefly study the stability of the model against small perturbations in sub-horizon scale around FRW background. 
In order to check the stability of the theory, let us perturb the background flat FRW metric as follows
\begin{equation}
ds^2=[1+\Upsilon (t, \textbf{x})]dt^2-a(t)^2[1+\Psi(t,\textbf{x})](dx^2+dy^2+dz^2)
\end{equation}
where $\Upsilon (t, \textbf{x})$ and $\Psi(t,\textbf{x})$ are small perturbations. Furthermore, for other fields we show the background and perturbed quantities with $0$ and $1$ subscripts, respectively. In this way, the corresponding perturbations in the vector field $\phi_{\alpha}$, and the scalar fields $\Phi$ and $\mu$ are respectively
 \begin{equation}
 \begin{split}
  & \bigr(\phi_t,\phi_x,\phi_y,\phi_z \bigr)=\bigr(\phi_{1t}(t,\textbf{x}), \phi_{1x}(t,\textbf{x}), \phi_{1y}(t,\textbf{x}), \phi_{1z}(t,\textbf{x})\bigr)
\\&\Phi=\Phi_{0}(t)+\Phi_1(t,\textbf{x})\\&\mu = \mu_0(t) + \mu_1(t,\textbf{x})
 \label{intro}
 \end{split}
 \end{equation}
It is necessary to mention that the background value of the vector field is zero in our case. One can easily verify this point using the field equation of the vector field, for example see \cite{Roshan:2014mqa}. Before moving on to write the linearised field equations, it is convenient to consider the perturbations as Fourier modes,  see e.g. \cite{lucasbook} for more details. Now using these assumptions and substituting the perturbations into the field equations, one may straightforwardly find the linearised field equations. In this case the field equations of $\Phi$ and $\mu$ up to the first order of perturbation are written as
\begin{equation}
 \ddot{\Phi}_1 + 3 H \dot{\Phi}_1 +\Big(\frac{2 \Lambda}{16 \pi - 3} +\frac{k^2}{a^2} \Big) \Phi_1=\Big( 3 H \Upsilon + \frac12 \dot{\Upsilon} - \frac{3}{2}\dot{\Psi} \Big) \dot{\Phi}_0 + \Upsilon \ddot{\Phi}_0
 \label{pphi}
\end{equation}
\begin{equation}
\ddot{\mu}_1 + \Big( 3H + \frac{\dot{\Phi}_0}{\Phi_0} - \frac{2 \dot{\mu}_0}{\mu_0} \Big) \dot{\mu}_1 + \frac{k^2}{a^2} \mu_1= F_1 \, \dot{\mu}_0 - \Big( \frac{\mu_1}{\mu_0^2} + \frac{\Upsilon}{\mu_0}\Big)\dot{\mu}_0^2 + \Upsilon \ddot{\mu}_0
\label{pmu}
\end{equation}
where the dot stands for derivative with respect to cosmic time $t$, and $F_1$ is defined as
\begin{equation}
F_1 = 3 H \Upsilon + \frac{\dot{\Phi}_0 \Phi_1}{\Phi_0^2}+ \frac{\dot{\Phi_0} }{\Phi_0} \Upsilon - \frac{\dot{\Phi_1}}{\Phi_0}+\frac12 \dot{\Upsilon} - \frac{3}{2}\dot{\Psi}.  
\end{equation}

On the other hand, the $ij$ component lead to the following algebraic relation, often called anisotropic stress, between metric perturbations and $\Phi_1$
\begin{equation}
\Psi + \Upsilon = -\frac{2 \Phi_1 }{\Phi_0}
\label{aniso}
\end{equation}

In the case of sub-horizon perturbations, the $\frac{k^2}{a^2}$ terms are dominant. Therefore, provided that for the background fields, denoted collectively as $Q$, $ \mid\dot{Q}_0\mid \lesssim \mid Q_0 H\mid $ and $\mid\ddot{Q}_0\mid \lesssim \mid Q_0 H^2\mid$, one can rewrite the equations (\ref{pphi}) and (\ref{pmu}) as
\begin{eqnarray}
& \ddot{\Phi}_1 + 3 H \dot{\Phi}_1 +\bigr(M^2 +\frac{3 k^2}{a^2} \bigr) \Phi_1 \simeq 0 \\
 & \ddot{\mu}_1 + 3H \dot{\mu_1} + \bigr(\frac{k^2}{a^2} \bigr)\mu_1\simeq 0
\label{pphiz}
\end{eqnarray}
where $M^2 \equiv \frac{2 \Lambda}{16 \pi - 3}>0$, which satisfies the necessary condition for the stability. On the other hand $\mu$ appears as a massless scalar field in this approximation.  Therefore one can conclude that there is no tachyonic instability in the sub-horizon limit of MOG.


\begin{thebibliography}{99}
\bibitem{moffat2}
  J. W. Moffat, JCAP
0603
, 004 (2006).

 \bibitem{Milgrom:1983ca} 
  M.~Milgrom,
  Astrophys.\ J.\  {\bf 270}  (1983) 365 .
  
      \bibitem{Bekenstein:2004ne} 
  J.~D.~Bekenstein,
  Phys.\ Rev.\ D {\bf 70} (2004)  083509 .

    \bibitem{Moffat:2013sja} 
  J.~W.~Moffat and S.~Rahvar,
  Mon.\ Not.\ Roy.\ Astron.\ Soc.\  {\bf 436}  (2013) 1439.
  
    \bibitem{Moffat:2014pia} 
  J.~W.~Moffat and V.~T.~Toth,
  Phys.\ Rev.\ D {\bf 91} (2015)  043004
  
  \bibitem{Moffat:2013uaa} 
  J. ~W.  ~Moffat and S.~Rahvar,
  Mon.\ Not.\ Roy.\ Astron.\ Soc.\  {\bf 441}  (2014) 3724
    \bibitem{Brownstein:2005dr} 
  J.~R.~Brownstein and J.~W.~Moffat,
  Mon.\ Not.\ Roy.\ Astron.\ Soc.\  {\bf 367}  (2006)  527
  
\bibitem{Ghafourian:2017wfr} 
  N.~Ghafourian and M.~Roshan,   Mon.\ Not.\ Roy.\ Astron.\ Soc.\  {\bf 468}  (2017) 4450; arXiv:1703.02718


\bibitem{Roshan:2016ygw} 
  M.~Roshan, S.~Abbassi and H.~G.~Khosroshahi,
  Astrophys.\ J.\  {\bf 832}, (2016) 201, arXiv:1610.01286.
  
  \bibitem{ra}
  M. Roshan and S. Abbassi, Phys. Rev. D {\bf90},
(2014) 044010,  arXiv:1407.6431.

  
  \bibitem{Roshan:2015gra}
  M.~Roshan and S.~Abbassi,
  Astrophys.\ J.\  {\bf 802} (2015), arXiv:1501.04715.
\bibitem{mn}
  M.~Roshan,
  Phys.\ Rev.\ D {\bf 87} (2013)  044005, arXiv:1210.3136.
\bibitem{Israel:2016qsf} 
  N.~S.~Israel and J.~W.~Moffat,
  arXiv:1606.09128 [astro-ph.CO].


   



 
  
 
  


  \bibitem{Roshan:2015uta}
  M.~Roshan,
  Eur.\ Phys.\ J.\ C {\bf 75} (2015), arXiv:1508.04243.
  
  
  \bibitem{Moffat:2014bfa} 
  J. ~W. ~Moffat, arXiv : 1409.0853
  
  \bibitem{Moffat:2015bdaf}
  J.~W.~Moffat,
  arXiv:1510.07037 [astro-ph.CO];
  J.~R.~Mureika, J.~W.~Moffat and M.~Faizal,
  Phys.\ Lett.\ B {\bf 757} (2016) 528;
  J.~W.~Moffat,
  Eur.\ Phys.\ J.\ C {\bf 75}, (2015) 130;  F.~Shojai, S.~Cheraghchi and H.~Bouzari Nezhad,
  Phys.\ Lett.\ B {\bf 770}, 43 (2017).
  
    
  
  
  
  
  
   
  \bibitem{Jamali:2016zww} 
  S.~Jamali and M.~Roshan,
  Eur.\ Phys.\ J.\ C {\bf 76}  (2016) 490, arXiv:1608.06251.  
  
  \bibitem{Amendola:1999qq} 
  L.~Amendola,
  Phys.\ Rev.\ D {\bf 60} (1999) 043501
  
   \bibitem{Amendola:2006kh} 
  L.~Amendola, D.~Polarski and S.~Tsujikawa,
  Phys.\ Rev.\ Lett.\  {\bf 98} (2007) 131302
  \bibitem{Amendola:2007nt} 
  L.~Amendola and S.~Tsujikawa,
  Phys.\ Lett.\ B {\bf 660} (2008) 125
  
  \bibitem{Moffat:2007nj} 
  J.~W.~Moffat and V.~T.~Toth,
  Class.\ Quant.\ Grav.\  {\bf 26} (2009) 085002
  
  \bibitem{Moffat:2015bda} 
  J.~W.~Moffat,
  arXiv:1510.07037 [astro-ph.CO].
  
\bibitem{Copeland:2006wr} 
  E.~J.~Copeland, M.~Sami and S.~Tsujikawa,
  Int.\ J.\ Mod.\ Phys.\ D {\bf 15} (2006) 1753
  
       \bibitem{Amendola:2006we}
  L.~Amendola, R.~Gannouji, D.~Polarski and S.~Tsujikawa,
  Phys.\ Rev.\ D {\bf 75} (2007) 083504
  \bibitem{haghi} 
  H.~Haghi and V.~Amiri,
  Mon.\ Not.\ Roy.\ Astron.\ Soc.\  {\bf 463} (2016) 1944
\bibitem{n2} Private communication with John Moffat, Victor Thoth and Martin Green. 

   \bibitem{Riess:2004} 
 P.~A.~R.~Ade {\it et al.} [Planck Collaboration],
  Astron.\ Astrophys.\  {\bf 594} (2016) A13
  
 \bibitem{Roshan:2014mqa} 
   M.~Roshan and S.~Abbassi,
   Phys.\ Rev.\ D {\bf 90}, no. 4, 044010 (2014) 

  \bibitem{lucasbook} 
  L. Amendola and S . Tsujikawa, Dark Energy
  Theory and Observations, Cambridge University Press, (2010), sec. (11).


\end{thebibliography}
\end{document}